\UseRawInputEncoding
\documentclass[a4paper,10pt,oneside]{article}
\usepackage[utf8]{inputenc}
\usepackage{icad2023,amsmath,epsfig,times,url,hyperref}

\usepackage[T1]{fontenc}
\usepackage{tabularx, booktabs}
\usepackage{chemformula}

\hypersetup{
    colorlinks,
    citecolor=darkgray,
    linkcolor=darkgray,
    urlcolor=darkgray
}

\title{Reef Elegy: An Auditory Display of Hawaii's 2019 Coral Bleaching Data}

\name{Stefano Kalonaris} 
\address{Music Information Intelligence Team \\ 
Center for Advanced Intelligence Project (AIP)\\ 
RIKEN, Japan \\ 
{\tt stefano.kalonaris@riken.jp}} 

\begin{document}
\ninept
\maketitle
\begin{sloppy}
    \begin{abstract}
    This paper describes an auditory display of Hawaii's 2019 coral bleaching data via means of spatial audio and parameter mapping methods. Selected data fields spanning 78 days are mapped to sound surrogates of coral reefs' natural soundscapes, which are progressively altered in their constituent elements as the corresponding coral locations undergo bleaching. For some of these elements, this process outlines a trajectory from a dense to a sparser, reduced soundscape, while for others it translates moving away from harmonic tones and towards complex spectra. 
    This experiment is accompanied by a short evaluation study to contextualize it in an established aesthetic perspective space and to probe its potential for public engagement in the discourse around climate change. 
    \end{abstract}
	
    \section{Introduction}
    \label{sec:intro}	
    Coral bleaching is a characteristic whitening of the visible surface caused by the expulsion of a microscopic unicellular algae called {\em zooxanthellae}, the photosynthetic pigments in corals.
	It happens in response to several factors such as low salinity, pollutants, and temperature stress, among others. While these have always been contributing causes, in the last few decades the role of humans in this process has been central. Recently, in fact, there has been an unprecedented increase in carbon dioxide and other significant greenhouse gases driven by fossil fuel combustion or agricultural and land-use sources (e.g., methane). The increase in greenhouse gas leads to an increase in air and sea temperature (the ``global warming''). Manifestations of these trends are changing sea levels and changing weather patterns (e.g., storms), both deleterious for coral reef ecosystems. Other anthropogenic factors that alter the natural ecosystemic balance of coral reefs are related to overfishing. This is because some fish (but also crabs, sea urchins, etc.) help maintaining coral health by eating reef macroalgae and preventing coral smothering and weakening. Therefore, overfishing for these species can result in a problematic increase in algae cover.
	Without further inquiry into the complex weaving of interactions contributing to coral bleaching, this paper looks at this phenomenon with respect to Hawaii. The first mass bleaching event in this area was reported in 2002 \cite{Aeby2003} with subsequent notable events in 2004, 2005, 2014, and 2015. Although not as severe as the 2014 and 2015 events, 2019's coral bleaching in Hawaii was still reason of concern and was documented extensively in a multi-institution initiative, leading to many online blogs and some publications \cite{10.1371/journal.pone.0269068}.
    Comparing pre- and post-bleaching via means of images (see Figure \ref{fig:coral_bleaching}) offers a stark warning and unequivocal measure of the urgency needed to implement drastic and long-lasting changes, both at socio-political and economic levels, but also in one's personal sphere (e.g., daily consumption habits, etc.). There are also tools for the visual assessment of coral bleaching that have been developed for the wider community, such as the Hawaiian Ko`a Card \cite{BahrEtAl2020}. The experiment presented in this paper sets out to investigate alternative representations of the same phenomenon for those who are less visually oriented and/or able, with the hope of contributing to raising awareness and calling for action to preserve and nurture marine ecosystems at large. To this end, some of the data available from a 2019 study (see Section \ref{subsec:data}) is used to auditorily display the process of coral bleaching by means of parameter mapping. Given the embryonic stage of this sonic experiment, the aims and goals are kept modest and exploratory rather than explanatory, as it does not try to uncover or discover correlations or causal relationships between data fields (e.g., pollutants, sea temperature, percentage of bleached coral populations, etc.) in the dataset. After a review of 1) relevant works that have treated similar topics and of 2) popular methods used for auditory display of data, this paper describes in detail the concept and the practical implementation of the author's experiment. Then, a small evaluation study is carried out and the results are discussed alongside future improvements in the context of public engagement on climate issues through sound.
	
	\begin{figure*}[!ht]
		\centering
		\begin{minipage}{.5\textwidth}
			\centering
			\includegraphics[width=0.75\linewidth]{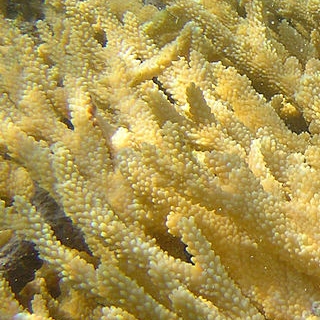}
		\end{minipage}%
		\begin{minipage}{0.5\textwidth}
			\centering
			\includegraphics[width=0.75\linewidth] {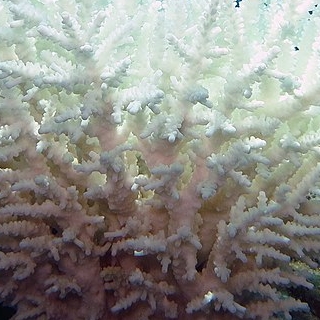}
		\end{minipage}
		\label{fig:coral_bleaching}
		\caption{Example of healthy (left) and bleached (right) Acropora coral, via Wikimedia Commons (common domain and CC BY-SA 4.0, respectively).}
	\end{figure*}
	
    \section{Related Work}
    \label{sec:related_work}
    Firstly, it is important to disambiguate the term {\em sonification}. Hereinafter, sonification will be used as a synonym of auditory display of data instead of the act of applying sound energy to agitate particles in a laboratory sample. It is necessary to clarify this point, because the alternate meaning is often found in marine biology literature and studies.
	
    Several marine phenomena/topics have been the object of interest in sonification endeavors, dealing with ocean conditions \cite{mead2021mayflower,10.1175/BAMS-D-15-00223.1,Sturm2002}, sea-surface water temperature \cite{10.2307/1577794}, or wideband hydroacoustic data \cite{MarcusWidmer2019}.
    As for coral reefs, NASA's {\em Coral Sea Sonification}\footnote{https://soundcloud.com/nasa/coral-sea-sonification} was created using spectral ocean reflectance data, and it is part of the {\em Sounds of the Sea}\footnote{https://www.nasa.gov/feature/goddard/2022/hear-sounds-of-the-sea-in-sonifications} initiative that focuses on auditory display.
	
    In recent years, multidisciplinary scientists and artists have also turned to rendering ecological data into sound in the context of art installations or more arts-oriented environments. Gilmurray termed this domain {\em Ecological Sound Art} \cite{gilmurray_2017,Gilmurray2016SoundingTA}. For example, maritime traffic has been addressed in sound art installations such as {\em Baltic Sea Radio}\footnote{https://var-mar.info/baltic-sea-radio/} by the artist duo Varvara \& Mar or the {\em nautical cycle} works of David Berezan (particularly {\em Sea Lantern}, 2017, and {\em Buoy}, 2011, both incorporating real-time data from sea buoys). 
	
    Regarding coral reef data in ecological sound art, Johnstone's {\em Coral Symphony} \cite{Johnstone2016CreatingAC} installation consisted of two separate sonification modules: one sonically displaying reef data from the Myrmidon Reef on the Great Barrier Reef, the other interactively mapping audience motion tracking data to seascapes obtained from undersea recordings, whereby the latter are injected, added and layered to a total resulting sound environment.
    In 2018, Lauren Jones and Eunjeong Stella Ko realized and presented {\em Hearing Seascapes}, a virtual-reality installation combining audiovisual data to generate endangered coral reef location-dependent sound.
	
    More recently, a dedicated concert titled {\em Sound as ocean memory: ecoacoustics, audification and sonification} \footnote{https://ccrma.stanford.edu/~brg/soniOM/april-9.html}, organized and held at CCRMA, Stanford, featured many compositions realized from undersea data, and two relating to coral, specifically. These were Mike Cassidy's {\em Coral Reef Sonification} (2021) which sonified data measuring gene expression in response to temperature fluctuations, and {\em choralCoral - 3 genomic \'{e}tudes of climate} (2021) by Tim Weaver, Steve Palumbi, and Jonathan Berger, that sonified Acropora coral species genomic sequences in relation to climate change and heat stress, in the Palau/South Pacific island region.
    Finally, in \cite{Spence2021LayersOM}, Heather R. Spence and Mark Ballora presented five sonifications and soundscape compositions based on marine data. Among these, {\em Reef Recall} comprises one (of six) movements (i.e., {\em Crustacean Chorus}) that relates to the characteristic shrimp-made crackling sonic tapestry of coral reefs. {\em Reef REM Ember}, instead, is another of those works, and layers processed (via means of transient detection, quantization, etc.) undersea recordings with input from live musicians. 

    \section{Sonification Approaches}
    \label{sec:sonification_model}
    There are distinct methods for auditory display of data, and the choice of one (or several) over the others is project-specific. For geodesic data, for example, {\em Audification} \cite{Worrall2019Audification,Dombois01usingaudification} is often used, but {\em Parameter Mapping Sonification} \cite{Worrall2019PMS}, hereinafter PMSon, remains arguably the indisputable default choice for most auditory display applications, allowing high customization in the sound design layers while maintaining (conditioned upon careful planning) intelligibility of the data. The ability to communicate or infer meaning in relation to the data or process of interest through sonification is a crucial factor in how successful an auditory display is judged. With this in mind, the authors of \cite{10.1109/JPROC.2004.825904} argue for a method called {\em Model Based Sonification}, in which users can ``explore'' a given phenomenon by interacting with a data model representation of it, by means of movement, for example. A more recent approach to auditory display of data goes by the name of {\em Wave Space Sonification} \cite{Hermann2018} whereby a scalar field is scanned along a data-driven trajectory.
    Finally, there exist other sonification methods, such as {\em Earcons} or {\em Auditory Icons}, that are used as auditory aids or sound placeholders for an event or action, for example, to alert users of a particular system's state.
	
    It appears that none of the relevant work cited in the previous section exploited the more recent paradigms of sonification. As for the experiment described in this paper, it is also designed using a more conventional approach (PMSon). Details of the model are presented in the next section.
	
    \section{Reef Elegy}
    \label{sec:reef_elegy}
    Most coral reefs are situated within the  boundary for $20^{\circ}$C isotherms. If one is in the habit of diving or snorkeling within this boundary, they probably have come across vast stretches of bleached or degraded reefs. It is not a sight for the faint of heart and, while one very much hopes for climate action and ecosystemic restoration, an {\em elegy} is an appropriate, if suggestive, term for a sound meditation on this topic. In fact, an elegy is a reflection or lament on death, normally intended in the form of poetry, but originally also included epitaphs (mournful songs).
	
    \subsection{Data}
    \label{subsec:data}
    The data was obtained from a two-month collection during 2019, on behalf of the Hawaii Coral Bleaching Collective (HCBC) \cite{hawaii2019} and comprises 517 clustered observations of 22 variables. Of these, only the mean percentage of living coral cover that was partially or fully bleached was used in the sonification, along with the mean geographic coordinates (in decimal degrees), the mean depth for the observation (ranging from $0.6$ to $29.8$ m), and the photosynthetically active radiation (PAR) values reported for each cluster.
    Figure \ref{fig:observations_map} shows location, depth and percentage of bleached coral population for these clusters. 
	
	\begin{figure}[!ht]
		\centering
		\includegraphics[trim={0 0 0 0},clip, width=\linewidth]{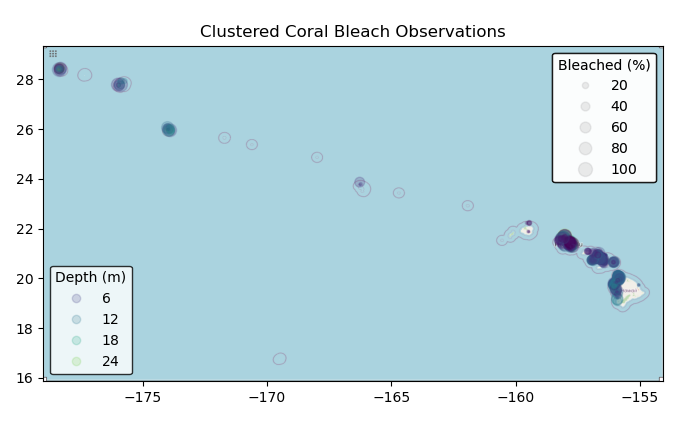}
		\caption{Data from the HCBC 2019 observations, relating to location, depth (m) and percentage of bleached coral population.}
		\label{fig:observations_map}
	\end{figure}
	
    However, the number of clustered observations proved to be CPU-intensive for the auditory display model (see Section \ref{subsec:spatialization}). Therefore, the data was further clustered based on the location, using the {\em ordering points to identify the clustering structure} (OPTICS) \cite{10.1145/304182.304187} algorithm (unsupervised) with a minimum number of samples $n=2$. This resulted in 176 clusters of location, bleaching, PAR, and depth averages. 
	
    \subsection{Spatialization}
    \label{subsec:spatialization}
    From this new dataset, a sonification dataframe was constructed as follows. The depth, PAR, and coral bleaching percentage were converted to the interval $[0, 1]$, while the mean longitude and latitude of each clustered data point was first re-scaled to the range of the original observations bounding box, then  converted to radians. 
    The reason for these steps lies in the fundamental idea behind {\em Reef Elegy}, which is to scatter the sonic representation of the observations on a 3D sound sphere. Thus, had the global Earth coordinate boundaries been kept, this would have resulted, perceptually, in a near-single point origin for the sound processes (corresponding to Hawaii's location on the planet). Regarding the use of radians instead of degrees, this is due to the requirements and specifications of the sound spatialization technique used, {\em ambisonics}. 
	
    Ambisonics \cite{Gerzon1973} is a full-sphere surround sound format that allows the encoding of a sound field independently from a given speaker configuration. The encoded sound field representation can then be decoded for playback on any speaker setup. The encoding can be done using variable spatial resolution, hereinafter referred to as {\em order}. For a given order $n$ a full-sphere system requires 
    $(n+1)^2$ channels. An ambisonic encoder takes a source signal $S$ and two parameters, the horizontal angle $\theta$ (or {\em azimuth}) and the elevation angle 
    $\phi$. It positions the source at the desired angle by distributing the signal over the ambisonic components. The convention for the coordinate system is that used for two-dimensional polar coordinates and three-dimensional cylindrical coordinates, but often times in physics and in the geographical spherical coordinate system $\theta$ is used for elevation (inclination) and $\phi$ for azimuth. For clarity, the following equivalences are stipulated and kept throughout: longitude or azimuth or $\theta$, in the interval $[-\pi, \pi]$, and latitude or elevation or $\phi$, in the interval $[-\pi/2, \pi/2]$.
    Figure \ref{fig:ambisphere} shows how the data points are encoded in an ambisonic sound field.
	
	\begin{figure}[!ht]
		\centering
		\includegraphics[trim={0 .75cm 0 .75cm}, clip, width=.85\linewidth]{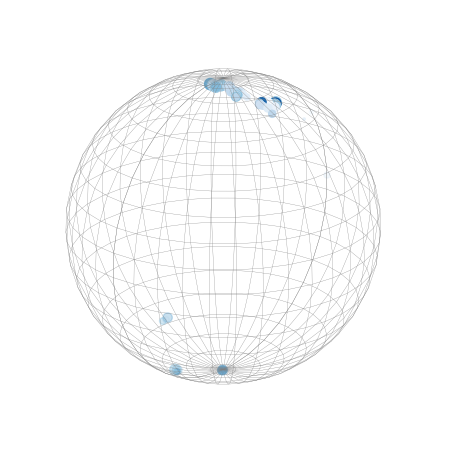}
		\caption{The re-clustered HCBC 2019 observations represented as sound sources on an ambisonic sphere.}
		\label{fig:ambisphere}
	\end{figure}
	
    For the spatialization as decribed above, the SC-HOA extension \cite{Grond2017HigherOA}, which provides wrapper classes of AmbiTools \cite{Lecomte2018} for the SuperCollider\footnote{https://supercollider.github.io} audio programming language, was used with an ambisonic order of $n=3$. This meant that the computational load for modeling the sound process with the original 517 sound sources posed challenges to the CPU used in this experiment. There needs to be a minimum of $517\times3$ {\em UGens} (building blocks of synth definitions, used to generate or process both audio and control signals) per sound source ($1$ encoder, $1$ decoder, and at least $1$ oscillator, for a minimal rendition). The original data was thus re-clustered as described in Section \ref{subsec:data}. 
	
    \subsection{Sonification}
    \label{subsec:sonification}	
    The sonification for {\em Reef Elegy} is inspired by the natural soundscape of coral reef environments. For the most part, these are dominated by the sounds of shrimp but there is also a component created by macroalgae's photosynthesis. The latter is normally difficult to detect due to the overwhelming shrimp snapping sound component, and because the two sound processes occupy approximately the same frequency band. However, the photosynthesis sound component has been hypothesized to be a sonic landmark of coral reefs under stress.
    Before proceeding with further details about the sonification model, the main assumptions adopted must be stated.
	
    First, that there exists some correlation between the presence of shrimp and the health of the coral reef, whereby the deterioration of the latter leads to a decreased population of the former. 
    While some \cite{Butler2017377} suggested that snapping shrimp frequency band sound level measurements can be useful as a metric to assess reef quality and biodiversity, others \cite{Freeman2016RapidlyOE} have shown little consistency in this regard.
    Nevertheless, the aforesaid relation is considered true for the sonification model, in this case. Moreover, it is assumed that the proportionality is real-time, so that there is a direct mapping between the percentage of bleached coral population and the density of the shrimp population, which can be maintained and dynamically (auditorily) displayed as time steps go by.
	
    Secondly, the relationship between macroalgae and coral is complex and involves many other factors, for example the role of herbivorous fish and sea urchins which is in turn affected by overfishing and sea pollution (see Section \ref{sec:intro}). An even remotely feasible modeling of this complex ecosystem at a sound level is beyond the scope of this paper's experiment. Here, it is simplistically assumed that the photosynthesis induced sound element is related to the PAR alone, overlooking all the contributing factors. 

    Third, both shrimp snapping and algae sound have peaks during diurnal times and valleys during the dark period of the day. These inter-day patterns are not considered in the experiment as, to compress 78 days into a time length suitable for listening, one day ends up being relatively short (one second, in the experiment described in Section \ref{subsec:experiment}). Therefore, the subtle waves of activity within it would not only be lost, but also, perhaps, mask the intra-day sound dynamics, which remain the focus of {\em Reef Elegy}.
	
    Two versions of the auditory display were produced, one done via sound synthesis, the other processing real undersea recordings. For simplicity, these will be referred to as {\em AD1} and {\em AD2}, respectively.
    With the 176 re-clustered data points, and according to the mapping strategies outlined so far, each sonification required an average of 1672 UGens.
	
    \vspace{.2cm}
    In both versions, and working with the aforesaid assumptions, two layers of sound were implemented: crackles and bubbles. 

    \subsubsection{Crackles}
    \label{subsubsec:crackles}
    Shrimp snap ``via the collapse of a cavitation bubble during rapid claw closure, generating broadband signals (up to 200 kHz), typically peaking between 2 and 20 kHz" \cite[p. 598]{Lillis2018}. 
	
    As a heuristic value, the minimum shrimp density estimations reported in \cite{Butler2017377} were chosen, for both degraded and healthy sites. These values originally referred to a single hydrophone coverage area, therefore they were scaled by three orders of magnitude considering that 1) the sonification represented many locations simultaneously, and 2) it would have been difficult to auditorily and perceptually appreciate the difference in density resulting from the bleaching data mapping. A shrimp density interval of [0.023, 0.471] was thus obtained and inversely mapped to the coral bleaching percentage using a linear to exponential mapping.
	
    For each day (a time step $t$) in the dataset, each sound source's crackles synth is passed the value of the coral bleach percentage for that cluster for that day as the density parameter. 
    Therefore, each cluster gradually reaches its percentage of coral bleach over the course of the sonification. 
    The gain value for each cluster's crackles synth is multiplied by a coefficient that accounts for the mean depth of the cluster, so that observations that were taken at higher depths get slightly boosted.  
	
    For AD1, the shrimp snapping/crackling was obtained using a UGen which generates random impulses according to the density parameter, whereas for AD2, the density parameter controlled the trigger rate of a sample-based granular synthesis UGen.  
	
    \subsubsection{Bubbles}
    \label{subsubsec:bubbles}
    As a result of algae photosynthesis, oxygen-containing bubbles are formed and subsequently released. In detaching from the algae, bubbles create a short ``ping'' sound. This bubble release is not uniform and appears ``as an irregular pulse-train-like time series'' \cite[p. 3]{10.1371/journal.pone.0201766}. 
    Tank experiments \cite{10.1371/journal.pone.0201766} should be considered with caution when porting findings to real coral reef environments; however, ``the bubble production mechanism [\ldots] may be used as a general indicator of photosynthetic activity" \cite[p.4]{10.1371/journal.pone.0201766}. 
	
    To auditorily render this process, the PAR values from the coral bleaching dataset were mapped to different parameters in the two sonifications.
    For AD2, the amplitude contribution from the bubbles element increases linearly according to the PAR daily value for each cluster. In AD1, the same is true except that, instead of using audio samples, each cluster's bubbles component is a frequency modulation (FM) unit, acting as a partial in a resulting additive synthesis (all clusters). The carrier frequency for each cluster's bubbles synth is an integer multiple of a given fundamental frequency (fixed), whereas the modulator frequency is affected by the PAR value and so is the index of modulation. Therefore, a relation between a complex and wide frequency spectrum and unhealthy coral reef is thus established. Consequently, towards the end of the sonification, the bubbles element is further from a harmonic tone than it was at the beginning. 
    
    In both cases (AD1 and AD2), and similarly to the crackles, the amplitude of a cluster's bubbles component is further multiplied by a depth compensation factor.

    \section{Evaluation}
    \label{sec:evaluation}
    Sonification, independent of the method or approach chosen, is inevitably going to involve a series of arbitrary decisions to establish formal relations between entities or properties in the data domain and their counterparts in the sound/music domain. Whether a sonification is more geared towards conveying information or is instead concerned with artistic goals is a topic that has informed many of the discussions in the auditory data display community. In \cite{VickersAndHogg2006}, a useful cartesian representation of the conceptual sonification space, subtended by an axis spanning from Ars Informatica to Ars Musica (the {\em Intentionality} dimension) and another spanning from Abstract to Concrete (the {\em Indexicality} dimension), is  put forward. This model, hereinafter the Aesthetic Perspective Space (APS), can help to contextualize a given project and to determine the conceptual position it occupies in the vast horizon of sonification. 
    More recently, the authors of \cite{Lindborg2023ClimateDS} formalized the notion of APS via an extensive study of $32$ sonification works relating to climate change. Inspired by the procedure outlined there and borrowing some of its metrics, a short evaluation of {\em Reef Elegy} was carried out.
	
    \subsection{Experiment}
    \label{subsec:experiment} 
    The study was anonymous and conducted online using a commercial but free survey tool part of a popular office suite offering. A total of $14$ participants completed the study, with an average completion time of $10$ min. Of the respondents, seven had variable degrees of familiarity with sonification (hereinafter, {\em initiated}) and seven had never heard of it before the study (referred to as {\em uninitiated}, from now on). A preliminary discriminative listening test asked everyone to evaluate two short ($3$ s) audio files. One was a recording of shrimp snapping/crackling, the other a synthesized abstraction of the natural process. Respondents simply had to state whether they thought the audio contained concrete or abstract sounds. Although, perhaps not surprisingly, the initiated did better in correctly classifying the two recordings (see Table \ref{tab:dl_task}), this was more of a tuning task, to get the participants' attention. Moreover, the different level of discernment did not seem to affect the statistical outcome of subsequent tests.
	
	\begin{table}[!ht]
		\centering
        \begin{tabularx}{.45\textwidth}{*{4}{X}}
		{} &  precision &  recall &  f-score \\
		\midrule
		Initiated   &      \textbf{0.889} &   \textbf{0.857} &    \textbf{0.854} \\
		Uninitiated &      0.733 &   0.714 &    0.708 \\
	\end{tabularx}
	\caption{Precision, recall, and f-score for the discriminative listening task.}
	\label{tab:dl_task}
	\end{table}

    Following this task, a brief description of {\em Reef Elegy} was given, to provide context, motivation, and essential information about the sonification project. Then, two main tasks were presented to the respondents, one relating to {\em Reef Elegy}'s position with respect to APS, and another about some qualitative characteristics of this sonification project. Both renditions of {\em Reef Elegy} were used for the experiment, with a time step $t = 1\mathrm{~s}$ and a fade out tail of $5$ s, for a total of $83$ s each. This choice was deemed appropriate for a listening task/study, with in mind to minimize the time load for the participants. 
	
    \subsubsection{APS}\label{subsec:aps}
    The APS questionnaire comprising eight Likert-items used in \cite{Lindborg2023ClimateDS} to span the APS was replicated here (for each sonification). Ratings used a seven-level response scale ranging from ``Strongly disagree'' to ``Strongly agree'' with ``Neutral'' in the middle. Variable name aliases for the items were also kept, to facilitate future comparisons of {\em Reef Elegy} with existing and already analyzed sonification works. A rapid visual inspection comparing results between the two groups suggested no significant differences. 
    Nevertheless, formal tests were carried out.
	
    First, the items' validity was tested via inter-rater agreement using Cronbach's alpha coefficient, yielding good results (shown in Table \ref{tab:ira_ad}).

	\begin{table}[!ht]
		\centering
        \begin{tabularx}{.45\textwidth}{*{4}{X}}
			{} &    AD1 &    AD2 &  ALL \\
			\midrule
			Both        &  0.759 &  0.777 &    0.881 \\
			Initiated   &  \textbf{0.794} &  0.786 &    \textbf{0.887} \\
			Uninitiated &  0.745 &  \textbf{0.788} &    0.884 \\
		\end{tabularx}
		\caption{Inter-rater agreement (Cronbach's $\alpha$) for the APS responses.}
		\label{tab:ira_ad}
	\end{table}

    Since the Likert-items were carefully designed to investigate the same construct and were fully Likert scoring-compliant (e.g., response levels were anchored with verbal labels which connote more-or-less evenly spaced gradations, bivalent and symmetrical about a neutral middle, arranged horizontally with equal spacing, etc.), it was thus reasonable to consider the recoded response levels (to consecutive positive integers) as intervals. This allowed to combine the Likert items into Likert-scales, and to apply parametric tests. Using mean as the aggregating operator, the resulting Likert-scales for the two groups were checked using an independent two-sample T-test with equal sample size. For both sonifications this yielded a $p > 0.05$, failing to reject the null hypothesis.
	
    To calculate the Intentionality and Indexicality values from the responses, the formulas used in \cite{Lindborg2023ClimateDS} were applied. Figure \ref{fig:aps} shows the perceived locations of the two sonifications on the APS, for the two groups. This further confirms that, regardless of the level of familiarity with auditory data display, each version occupied a similar conceptual space. Moreover, the two sonifications scored similarly, despite one being obtained using real-world undersea recordings and despite the reasonable ability across both groups to discriminate between these sounds and their synthesized simulations/models. Thus, applying (granular) synthesis to natural soundscapes and naturalizing synthetic sounds (through modeling) had the effect of pulling the two sonic outputs closer to one another. 

    \begin{figure*}[ht!]
		\centering
		\begin{minipage}{.48\textwidth}
			\centering
			\includegraphics[trim={0 .7cm 0 0}, clip, width=0.97\linewidth]{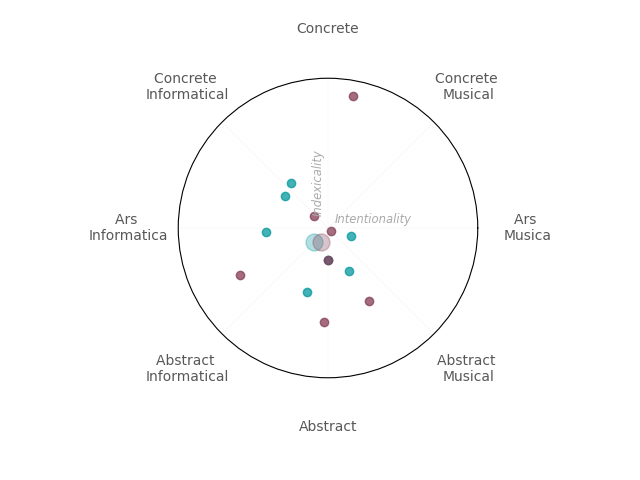}
		\end{minipage}%
		\begin{minipage}{0.48\textwidth}
			\centering
			\includegraphics[width=0.97\linewidth] {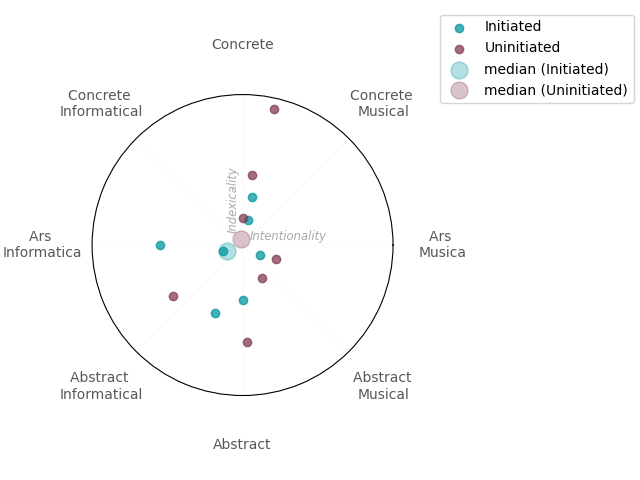}
		\end{minipage}
		\caption{Perceived location on the APS for AD1 (left) and AD2 (right).}
  		\label{fig:aps}
	\end{figure*}
	 
    \subsubsection{Qualitative Characteristics}
    \label{subsec:qchar}
    As for the qualitative characteristics items, the original list in \cite{Lindborg2023ClimateDS} was aimed at sonification experts who evaluated finished works that had been presented, performed, or published at the time of evaluation. Therefore, that study presupposed a considerable amount of information on the object of the evaluation.
    Considering that the study presented here refers to the prepublication stage of a sonification endeavor, not necessarily aimed at domain experts, and under further anonymity constraints due to the review process, most of the aforementioned scales seemed problematic to this end. Notwithstanding, five out of the original 25 items were kept to gather insight. The corresponding original alias/convenience variable names were also kept, and items were evaluated on a seven-level rating scale ranging from ``Extremely little'' to ``Extremely much'' with ``Average'' in the middle.  
	
	\begin{figure*}[ht!]
		\centering
		\begin{minipage}{.48\textwidth}
			\centering
			\includegraphics[width=0.86\linewidth]{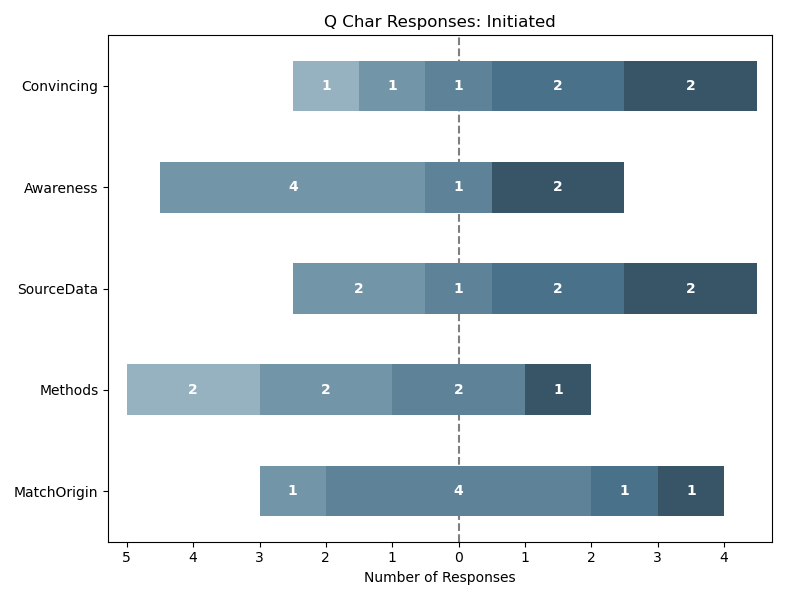}
		\end{minipage}%
		\begin{minipage}{0.48\textwidth}
			\centering
			\includegraphics[width=0.97\linewidth] {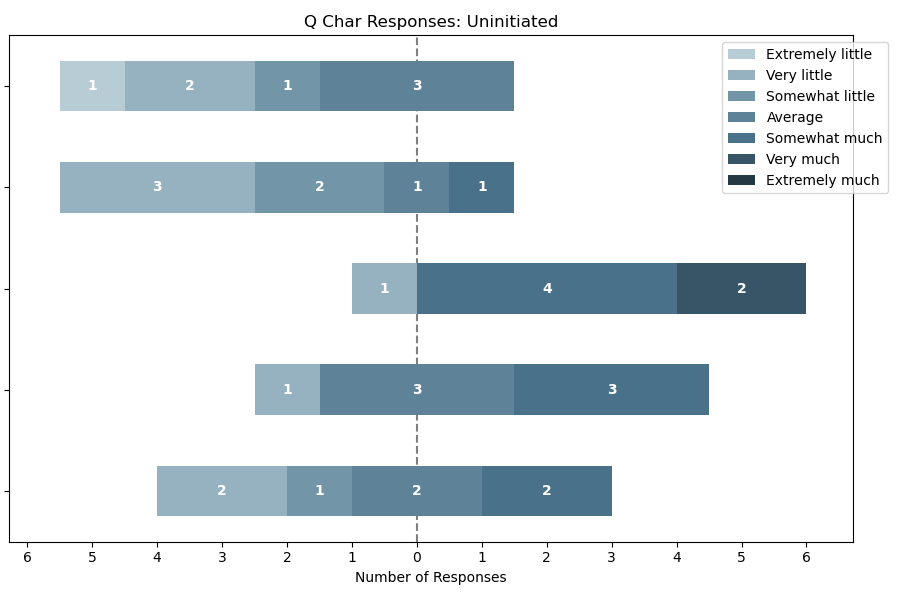}
		\end{minipage}
		\caption{Responses regarding qualitative characteristics, for all sonifications, for both groups: initiated (left) and uninitiated (right).}
  		\label{fig:qc_likert}
	\end{figure*}
	
    The first item asked how convincing the sonifications were in communicating coral bleaching, while the second was related to the perceived potential of the sonifications in raising awareness or thoughts about climate change. These were followed by questions on the specificity of the information on source data, on the level of detail about the sonification methods and, lastly, on how well the sonifications matched the original phenomenon. 
    These questions were asked only once, for both sonifications to be considered as a standalone project.
    Figure \ref{fig:qc_likert} shows the responses per group for these items.
	
    It would appear that the most contended item was the first (variable alias name: Convincing), with the initiated group leaning more towards a positive response (four people responded with levels above the neutral middle) when compared to the uninitiated (no positive leveled responses). 
    On the contrary, there seemed to be consensus on the modest if poor potential of the sonifications to raise awareness,  on an appropriate level of information about the source data, and on the satisfactory representation of the original phenomenon. Finally, respondents familiar with sonification seemed to have found the details regarding the mapping methods somewhat lacking.
    Since the items did not inquire about the same construct, further investigation did not follow the same procedure adopted earlier for the APS tasks. Instead, non-parametric tests (i.e., pairwise Mann-Whitney U) were conducted, which all failed to reject the null hypothesis.

    \smallskip
    The study did not provide participants with optional fields for commentary. However, one of the respondents reached out to share their thoughts on the sonifications, which are reported below verbatim.

	\begin{quotation}
		Although it would have been much easier to convey and faster to consume the message of coral bleaching by just displaying a conventional visualization, sonification adds a lot of value emotionally and engages the listener more. The temporal aspect of how the bleaching process is unfolding is best conveyed by sonification which requires time to consume the information and learn about a phenomena. Also, the sound of unhealthy coral reef has this ominous darker quality to it, and when it slowly overtakes the rhythmic joyful cracking sound of the shrimp, listener can emotionally connect with the process of coral reefs suffering bleaching, shrimp being swept away by a wave of dull hissing sound signifying the crisis unfolding. (Participant, private email communication)
	\end{quotation}
		
    \section{Conclusion}
    \label{sec:conclusion}
    This paper presented {\em Reef Elegy}, a parameter mapping sonification of Hawaii's 2019 coral bleaching data, evaluated by means of an online survey with participants of varying sonification skill level.
    The consensus that emerged seemingly positioned this work around the center of an established conceptual space for aesthetic perspective in sonification, and deemed the proposed auditory display effective in portraying the gradual loss of life and diversity in coral ecosystems. At the current prototypical stage of development, it did not prove to be sufficiently inspiring for the debate around climate change. 
	
    To this end, it is reasonable to envisage renditions of {\em Reef Elegy} where both the time span and the complexity of the mapping can be extended. 
    Adjusting and/or replacing some of the modular elements in the source code would allow to achieve results more akin to sound art, regarding both sonic content and time scales (one can think of some slowly evolving works by Eliane Radigue, such as {\em Transamorem - Transmortem}, for example). Furthermore, because of the ambisonic encoding of the sonification process, rendering to multichannel speaker setup would be seemingly automatic. A domain shift from the current small experiment towards some more immersive future edition would arguably benefit the aims and goals of this sonic endeavor. Given the increasing occurrence and scale of mass bleaching events, it is hoped that continuing to offer alternative, improved, and more engaging representations of such critical phenomena might help sensibilize audiences and inspire structural, sustained action and change in policies, behavior, and habits.

    \section{Acknowledgment}
    \label{sec:acknowledgment}

    The author thanks Georgios Diapoulis for standing in for him during the conference presentation. To avoid an unnecessary carbon footprint (flying from Tokyo to Stockholm and back generates about 1,475 kg \ch{CO2}) and in line with this paper's climate responsibility pledge, the author did not attend in person.
    
    \bibliographystyle{IEEEtran}
    \bibliography{bibliography}
    
    \appendix
    \section{Supplementary Material}
    Supplementary material regarding Section \ref{sec:evaluation}, including the two sonifications, is available online at
    \url{https://doi.org/10.5281/zenodo.7879979}

    \end{sloppy}
    
\end{document}